\newcommand{\be}{\begin{equation}}\newcommand{\ee}{\end{equation}}
\newcommand{\bea}{\begin{eqnarray}}\newcommand{\eea}{\end{eqnarray}}
\newcommand{\nn}{\nonumber}\newcommand{\p}[1]{(\ref{#1})}
\newcommand{\lb}[1]{\label{#1}}
\newcommand\s{\scriptscriptstyle}

\newcommand\q{\quad}
\newcommand\qq{\quad\quad}

\newcommand\cB{{\cal B}}

\newcommand\cF{{\cal F}}
\newcommand\cG{{\cal G}}

\newcommand\cP{{\cal P}}

\newcommand\tpa{\theta^{+\alpha}}

\newcommand\tma{\theta^{-\alpha}}

\newcommand\btpa{\bar{\theta}^{+\dot{\alpha}}}

\newcommand\btma{\bar{\theta}^{-\dot{\alpha}}}

\newcommand\tp{\theta^+}

\newcommand\btp{\bar\theta^+}

\def\a{\alpha}
\def\da{{\dot\alpha}}
\def\b{\beta}
\def\db{{\dot\beta}}
\def\g{\gamma}
\def\dg{{\dot\gamma}}
\def\d{\delta}

\def\eps{\epsilon}
\def\bep{{\bar\epsilon}}          
\def\ve{\varepsilon}

\def\bph{{\bar\phi}}

\def\l{\lambda}

\def\o{\omega}

 \def\th{\theta}  \def\bt{\bar\theta}
\def\vt{\vartheta}
\def\r{\rho}

\def\si{\sigma}  
\def\bs{\bar\sigma}

\def\D{\Delta}

\def\J{\Psi}
\def\bJ{\bar\Psi}
\def\L{\Lambda}

\def\pa{\partial}

\newcommand\ada{{\alpha\dot{\alpha}}}

\newcommand\bdb{{\beta\dot{\beta}}}
\newcommand\bda{{\beta\dot{\alpha}}}

\newcommand\ab{{\alpha\beta}}

\newcommand\pada{\partial_{\alpha\dot{\alpha}}}

\newcommand\A{{\s A}}

\newcommand\R{{\s R}}
\newcommand\M{{\s M}}
\newcommand\sL{{\s L}}
\newcommand\N{{\s N}}
\newcommand\W{{\s W}}
\newcommand\Z{{\s Z}}

\newcommand\hx{{\hat{x}}}

\newcommand\hf{{\hat{f}}}
\newcommand\hg{{\hat{g}}}

\newcommand\hd{{\hat\delta}}
\newcommand\hA{{\hat{A}}}
\newcommand\hB{{\hat{B}}}
\newcommand\hG{{\hat{G}}}
\newcommand\hM{{\hat{M}}}
\newcommand\hL{{\hat{L}}}

\newcommand\hX{{\hat{X}}}
\newcommand{\pp}{{\s ++}}

\newcommand{\Dp}{D^{\pp}}

\newcommand{\dpp}{\partial^{\pp}}

\newcommand{\Vp}{V^\pp}

\newcommand{\Dpa}{D^+_\alpha}
\newcommand{\Dma}{D^-_\alpha}

\newcommand{\bDpa}{\bar{D}^+_{\dot{\alpha}}}

\newcommand{\bDma}{\bar{D}^-_{\dot{\alpha}}}

\def\sfrac#1#2{{\textstyle\frac{#1}{#2}}}

\def\ii{\mbox{i}}

\documentclass[12pt]{article}
\usepackage{amsmath,amssymb}

\renewcommand{\thefootnote}{\fnsymbol{footnote}}

\begin{document}
\begin{center}
{\Large\bf  DEFORMATIONS OF EUCLIDEAN SUPERSYMMETRIES}\\
\vspace{1cm}

{\large\bf
  B.M. Zupnik\footnote{Joint Institute for Nuclear Research, Dubna,
  Moscow region, Russia; E-mail: zupnik@theor.jinr.ru}}
\vspace{1cm}

\end{center}

\begin{abstract}
We consider quantum supergroups that arise in non-anticommutative
deformations of  $N{=}(\frac12,\frac12)$ and $N{=}(1,1)$ four-dimensional
Euclidean supersymmetric theories. Twist operators in the corresponding
deformed algebras of superfields contain left spinor generators.
We show that non-anticommutative  $\star$-products of superfields
transform covariantly in the deformed supersymmetries. This covariance
guarantees the invariance of deformed superfield actions of models
involving $\star$-products of superfields.
\end{abstract}

Key words: Supersymmetry, superspace, deformation, twist

\renewcommand{\thefootnote}{\arabic{footnote}}
\setcounter{footnote}0
\setcounter{equation}0
\section{Introduction }

The simplest type of the space-time noncommutativity is connected with the
 relation  
\be
C^{mn}_\star\equiv\hx^m\star \hx^n-\hx^n\star \hx^m-i\vt^{mn}=0 \lb{A}
\ee
for the coordinate operators $\hx^m$, where $\vt^{mn}$ ($m, n=0, 1, 2, 3$) is 
some constant tensor specifying
the deformation of the commutative four-dimensional coordinates $x^m$ (see, e.g. 
 \cite{Rev1,SW}). The noncommutative algebra of fields $\hf(\hx)$ on this 
deformed space-time is formally analogous to the Weyl algebra on the quantized 
phase space $\hx, \hat{p},\q [\hat{p},\hx]=i\hbar$.
The noncommutative algebra $A_\star$ is defined as the algebra of formal 
polynomials in $\hx^m$ factored over quadratic relation \p{A}. The Weyl ordering 
of the operator field  involves a decomposition in terms of
completely symmetrized monomials $\hx^{(m_1}\star\ldots\star\hx^{m_n)}$
\be
 \hf(\hx)=\sum\limits_{n=0}^\infty c_{m_1\ldots m_n}\hx^{(m_1}\star\ldots\star
 \hx^{m_n)},
\ee
where $c_{m_1\ldots m_n}$ are numerical coefficients that are symmetric in 
$m_1,\ldots m_n$. We can use the correspondence of this ordered operator 
function and an ordinary smooth function $f(x)$ 
\be
w[\hf(\hx)]= f(x)=\sum\limits_{n=0}^\infty c_{m_1\ldots m_n}x^{m_1}\ldots 
x^{m_n},
\ee
which we  call the commutative image of $\hf(\hx)$. The map inverse  to
the operator representation is denoted by $w^{-1}[f(x)]=\hf(\hx)$.

A  realization of this  noncommutative algebra $A_\star$ that is popular in 
field theory can be defined on smooth  functions $f(x)$ and $ g(x)$ of commutative
coordinates using the following pseudolocal representation of the noncommutative 
product:
\bea
&&w[\hf\star\hg]=f\star g=f e^P g=fg+\sfrac{i}2\vt^{mn}\pa_mf\pa_ng-
\sfrac18\vt^{mn}\vt^{rs}\pa_m\pa_r f\pa_n\pa_sg+\ldots,\nn\\
&&\lb{star} fPg=\sfrac{i}2\vt^{mn}\pa_mf\pa_ng,
\eea
where $\pa_m=\pa/\pa x_m$. All products of the functions and their partial
derivatives in the right-hand side are commutative. The formula $(\hf\star\hg)(\hx)
=w^{-1}[(f\star g)(x)]$ allows constructing the ordered decomposition
of the noncommutative product of operator functions via the power series
of products of smooth functions. Basic commutational relation \p{A}
is identically satisfied in representation \p{star}
\be
 x^m\star x^n=x^m x^n+\sfrac{i}2\vt^{mn}. \lb{P}
\ee
Very many results of investigations in noncommutative field theory
are obtained exactly in this convenient field theory representation using
arbitrary-order derivatives of local fields, although all properties of
the theory can be reformulated in the operator representation.

Let us consider a simple $\vt$-noncommutative interaction of the real
scalar field $\phi(x)$ in the noncommutative algebra  $A_\star$
\be
S_\star=\int d^4x L_\star(\phi)=\sfrac12\int d^4x(\eta^{mn}\pa_m\phi\star
\pa_n\phi-M^2\phi\star\phi -\lambda \phi^4_\star),\lb{noncom}
\ee
where $M$ and $\lambda$ are the mass and coupling constant, respectively, 
$\eta^{mn}$ is the Minkowski space metric, and $\phi_\star^4=
\phi\star\phi\star\phi\star\phi$. In  representation \p{star}, quadratic
terms contain the standard undeformed free action and additional vanishing
integrals of total derivatives. The nonlinear interaction depends manifestly
on $\vt_{mn}$, and is therefore not invariant with respect to the standard 
Lorentz transformation. Despite this breaking of the Lorentz invariance, all 
models of the $\vt$-noncommutative field theory use the following selection rule:
{\it  Basic noncommutative (primary) fields  transform as representations
of the Poincar\'e group, and interactions of these fields are constructed
on the basis of algebra  $A_\star$}. With the $\star$-products being noncovariant
with respect to the Lorentz group, this rule has no simple interpretation
in the framework of usual symmetries.

As shown in \cite{Oe}-\cite{KoM}, the selection rule follows
from the invariance of the noncommutative field theory under
the $t$-deformed quantum Poincar\'e group involving the twist operator
\be
\cF=\exp(\sfrac{i}2\vt^{mn}\pa_m\otimes \pa_n)\lb{Ptwist}.
\ee
Note that other variants of quantum-group deformations of the
Poincar\'e group were previously been considered \cite{PW}-\cite{LRTN}. 
In particular, various forms of the Drinfeld twist operator \cite{Dr}
were widely used in these investigations. Nevertheless, the corresponding
field models have not been studied in as much  detail as the models based on 
relation \p{A}.

In sect. 2, we consider the $t$-deformed Lorentz transformations for
the $\star$-product of primary fields using local relations between
the differential operators on commutative and noncommutative algebras
in  \cite{We,ABDMSW}. 

Deformations of supersymmetric theories are characterized by a Poisson
bracket $APB$ on the superfields $A$ and $B$, where the operator $P$ is a
general quadratic form in terms of derivatives with respect to the even
and odd coordinates in the superspace \cite{FL,KPT}. Important classes
of  nilpotent $Q$-deformations of the Euclidean supersymmetries were
found in \cite{Se} for the case of $N{=}(\sfrac12,\sfrac12), D=4$ supersymmetry
and in \cite{ILZ,FS} for the $N{=}(1,1), D=4$ supersymmetry. The deformation
operators $P$ for these models are constructed from the left spinor 
generators of supersymmetries $Q$. They preserve superfield constraints
for the undeformed supersymmetry representations, for instance,
the chirality or Grassmann analyticity constraints. The $Q$-deformed
superfield theories \cite{Se}-\cite{ILZ2} use $\star$-products of undeformed
primary superfields in the pseudolocal representation.
Deformed theories are not invariant under  the action of generators
of the basic Euclidean supersymmetry that do not commute with  $P$.

The interpretation of deformed Euclidean supersymmetries in the framework
of the Hopf algebras was introduced in \cite{KS}. Section 3 is devoted to 
 discussing of the twist-deformed  $N{=}(\sfrac12,\sfrac12)$
supersymmetry. The non-anticommutative $\star$-product in the deformed 
superalgebra is defined on supercommutative superfields in the ordinary
superspace. Primary superfields of this model are transformed as representations
of $N{=}(\sfrac12,\sfrac12)$ supersymmetry  realized by the first-order
differential operators. The twist operator $\cF$ for the nilpotent $Q$-deformation
determines the coproduct in the $t$-deformed $N{=}(\sfrac12,\sfrac12)$
supersymmetry and also the $\star$-product in the corresponding
noncommutative algebra of superfields. We formulate the local covariance
principle in this algebra: primary superfields $A, B$ and their 
$\star$-product $A\star B$ have the same transformations in the $t$-deformed
supersymmetry. Generators of the deformed transformations in the operator
representation are uniquely defined  by the twist operator and the
corresponding undeformed supersymmetry generator. For instance,
the deformed generators of the right supersymmetry transformations contain
deformation constants and the second-order Grassmann derivatives  
in the operator representation, although the Lie superalgebra remains
undeformed. We show that the deformed superfield actions using
the $\star$-product preserve the invariance under the
$t$-deformed supersymmetry.

In sect. 4, we consider the twist operator defining the coproduct in
the $t$-deformed   $N{=}(1,1)$ supersymmetry  and the corresponding
$\star$-product in the non-anticommutative algebra of harmonic superfields. 
The deformed superfield actions constructed in the $N{=}(1,1)$ harmonic 
superspace \cite{FILSZ}-\cite{CILQ} partially violate the standard
supersymmetry, but these superfield actions are invariant under
 the transformations of the $t$-deformed $N{=}(1,1)$ supersymmetry.
It is notable that the quadratic superfield terms of the action also
preserve the ordinary supersymmetry.

It should be remarked that the quantum-group deformations of the supersymmetry
with more complex superfield geometry were previously considered \cite{KMLS},
but we do not discuss these models in our work. 
The $t$-deformed supersymmetries in the superfield theories were briefly
described in \cite{Zu}. 

\setcounter{equation}0
\section{Deformed Poincar\'e group }
We  review  the basic applications of the $t$-deformed Poincar\'e
symmetry \cite{Oe}-\cite{ABDMSW} in the noncommutative field theory.
The corresponding twist operator   $\cF =\exp(\cP)$ \p{Ptwist} acts on
tensor products of functions
\bea
&&\cF\circ f\otimes g =f\otimes g+\cP\circ f\otimes g+\sfrac12\cP^2\circ 
f\otimes g+\ldots,\nn\\
&& \cP\circ f\otimes g=\sfrac{i}{2}\vt^{mn}\pa_mf\otimes \pa_ng.\lb{twist}
\eea
The rigorous definition of  noncommutative product \p{star} is related
to the operator  $\cF$
\be
f\star g\equiv\mu_\star\circ f\otimes g=\mu\circ \cF\circ f\otimes g,\q 
\mu\circ f\otimes g=fg,
\lb{startw}
\ee
where the map $\mu$ defines the product in the commutative algebra of functions
$A(R^4)$, and the analogous map $\mu_\star=\mu\circ \cF$ defines the product
in the algebra $A_\star$. The twist operator is analogous to the pseudolocal
operator $e^P$ \p{star} in the field theory constructions, and the
tensor product can be treated as a nonlocal product of ordinary fields
in different points $f(x_1)g(x_2)$.
 
We consider a local representation of the generators of the Poincar\'e 
group
\bea
&&P_m=\pa_m,\q M_{mn}=x_n\pa_m-x_m\pa_n\lb{Pgen}
\eea
and the corresponding infinitesimal transformations of the scalar field
\bea
\d_c\phi=-c^mP_m\phi=-P_c\phi,\q\d_\o\phi=-\sfrac12\o^{mn}M_{mn}\phi=-M_\o\phi,
\eea
where $c_m, \o^{mn}$ are the infinitesimal parameters of translations and
Lorentz transformations. A finite transformation of the form  of the 
scalar function (active Poincar\'e transformation) can be represented as
\be
\phi^\prime(x)=\phi(\tilde{x})=e^{-M_\o}e^{-P_c}\phi(x),\q \tilde{x}^m=
e^{-M_\o}(x^m-c^m).
\lb{global}
\ee
The transformation operator $e^{-M_\o}e^{-P_c}$ belongs to the universal
enveloping bialgebra of functions of generators $U(P_m,M_{mn})$, where
the associative product of generators and coproduct $\D:~U\rightarrow 
U\otimes U$ are defined. The coproduct acts trivially on  unity and
the generators of  $U(P_m,M_{mn})$ 
\bea
&&\D(1)=1\otimes 1,\q\D(P_c)=P_c\otimes 1+1\otimes P_c,\q \D(M_\o)=M_\o\otimes 1
+1\otimes M_\o,\lb{triv}
\eea
and the action of $\D$ on functions of generators is  defined accordingly,
for instance, 
\bea
&&\D(M_\o M_\o)=\D(M_\o)\D(M_\o),\q \D(e^{-M_\o})=e^{-M_\o}\otimes e^{-M_\o}.
\nn
\eea
The coproduct gives the action of generators and their functions on
the tensor product of fields
\be
\D(M_\o)\circ f\otimes g=M_\o f\otimes g+f\otimes M_\o g.
\ee
Thus, the standard Leibniz rule for the infinitesimal Lorentz transformation
in the commutative algebra follows from the formula for $\D(M_\o)$ 
\be
\d_\o (fg)=\mu\circ\d_\o(f\otimes g)=(\d_\o f)g+f\d_\o g.
\ee
This rule is postulated in the commutative field theory, but deformations
of the coproduct and the corresponding transformation laws of the products
of functions are possible in the noncommutative case.

It is evident that the noncommutative product transforms noncovariantly in
the ordinary Lorentz group
\be
\d_\o(f\star g)=(\d_\o f)\star g+(f\star \d_\o g)\neq -M_\o(f\star g).
\ee

By definition, the $t$-deformation of the Poincar\'e group: $U(P_m,M_{mn})
\rightarrow U_t(P_m,M_{mn})$ does not change the Lie algebra of generators,
and we can therefore use the standard representation \p{Pgen} in $U_t$. The 
coproduct in $U_t(P_m,M_{mn})$ is deformed for generators $M_{mn}$
\bea
&&\D_t(P_c)=\exp(-\cP)\D(P_c)\exp(\cP)=\D(P_c),\nn\\
&&\D_t(M_\o)=\exp(-\cP)\D(M_\o)\exp(\cP)=\D(M_\o)\nn\\
&&+\sfrac{i}2\o^{mn}\vt_{ms} P_n\otimes P^s
-\sfrac{i}2\o^{mn}\vt_{rn} P^r\otimes P_m.
\lb{coprM}
\eea 

The deformed transformations of fields in the noncommutative algebra $A_\star$ 
was described in details in the recent works 
of the Munchen group \cite{We,ABDMSW}. These authors  constructed a map
between differential operators on the commutative and noncommutative
algebras of functions. Let $\xi=\xi^m(x)\pa_m$ be the first-order
differential operator on  $A(R^4)$ containing an arbitrary function
$\xi^m(x)$; for instance, the infinitesimal Lorentz transformation is
defined by the function $\xi^m=\o^{mn}x_n$. One can construct the corresponding
operator $\hX_\xi$ on the noncommutative algebra $A_\star$ satisfying
the simple relation
\bea
&&(\xi f)=\mu\circ (\xi^m\otimes\pa_m)(1\otimes f)=\mu_\star\exp(-\cP)
(\xi^m\otimes\pa_m)(1\otimes f)=\xi^m\star\pa_m f\nn\\
&&-\sfrac{i}2\vt^{rs}\pa_r\xi^m\star\pa_s\pa_m f
-\sfrac18\vt^{rs}\vt^{pq}\pa_r\pa_p\xi^m\star\pa_s\pa_q\pa_m f+O(\vt^3)
=(\hX_\xi\star f).\lb{diffmap}
\eea
Note that in the general case $\hX_\xi$ contains derivatives of arbitrary
orders and acts, by definition, on any noncommutative functions or their
commutative images.

The local generators $P_m$ and $M_{mn}$ \p{Pgen} acting on commutative images
$f(x)$ correspond to the following operators on the noncommutative functions
$\hf(\hx)$ in algebra $A_\star$:
\bea
&&\pa_m f=(\hat P_m\star f),\q (\hat P_m\star\hat f)=w^{-1}(\hat P_m\star f),
\nn\\
&& (M_{mn}f)=(\hat M_{mn}\star f),\q (\hat M_{mn}\star\hat f)=
w^{-1}(\hat M_{mn}\star f) .\lb{hatop}
\eea
We note that the translation generator  has the standard form, and the
generator of the Lorentz transformation in the operator representation
contains  second-order derivatives
\bea
&& \hat M_{mn}=x_n\star\pa_m-x_m\star\pa_n+\sfrac{i}2\eta_{mr}\vt^{rs}\pa_s
\pa_n-\sfrac{i}2\eta_{nr}\vt^{rs}\pa_s\pa_m,\nn\\
&&\hM_\o=\sfrac12\o^{mn}\hat M_{mn}=\o^{mn}x_n\star\pa_m+\sfrac{i}2\o^{mn}
\eta_{mr}\vt^{rs}\pa_s\pa_n.\lb{hM}
\eea
It is natural to use operators $\hat P_m=\pa_m$ and $\hat M_{mn}$ in the
operator representation of the noncommutative algebra $A_\star$ on functions
$\hf(\hx)$, for instance,
\bea
&&(\hat M_\o\star \hx^m\star\hx^n)=\eta_{sr}[\o^{ms}\hx^r\star\hx^n+\o^{ns}
\hx^m\star\hx^r-\sfrac{i}2\o^{ms}\vt^{rn}+\sfrac{i}2\o^{ns}\vt^{rm}].
\eea

It is easy to verify that the quantity $C^{mn}_\star$ \p{A} is covariant
with respect to the transformations of the deformed Lorentz group
\be
(\hat M_\o\star C^{mn}_\star)=\o^{ms}\eta_{sr} 
C^{rn}_\star+\o^{ns}\eta_{sr}C^{mr}_\star.
\ee
The anticommutator $A^{mn}_\star=\hx^m\star\,\hx^n+\hx^n\star\,\hx^m$  also 
transforms covariantly in $U_t(P_m,M_{mn})$.

As  follows from relations \p{hatop}, the operators $\hat P_m$ and $\hat M_{mn}$
form the Lie algebra (and the corresponding associative algebra) isomorphic
to the standard algebra with the generators $P_m$ and $M_{mn}$ 
\be
(\hat M_{mn}\star\hat M_{rs}\star \hf)=(M_{mn}M_{rs}f) ,\q (\hat P_s\star
\hat M_{mn}\star \hf)=(P_sM_{mn}f) .\lb{hatal}
\ee

Coproduct  \p{coprM} acts on  tensor products of functions as
\bea
&&-\sfrac12\o^{mn}\D_t(M_{mn})\circ f\otimes g=(\d_\o f)\otimes g+f\otimes
(\d_\o g)\nn\\
&&+\sfrac{i}2(\o^{mn}\vt_{ns}-\vt^{mn}\o_{ns})\pa_mf\otimes \pa^sg.\lb{dtens}
\eea
We apply the map $\mu_\star$ \p{startw} to this relation and obtain the
formula for the covariant action of the deformed Lorentz transformation
in the algebra $A_\star$
\bea
&&\hd_\o\star(f\star g)= -\sfrac12\o^{mn}\mu_\star\circ\D_t(M_{mn})\circ 
f\otimes g=(\d_\o f)\star g+f\star(\d_\o g)\nn\\
&&+\sfrac{i}2(\o^{mn}\vt_{ns}-\vt^{mn}\o_{ns})\pa_mf\star \pa^sg= 
-M_\o(f\star g).
\eea
The last relation can be verified using the pseudolocal formula 
\p{star}. Thus, the noncommutative product of scalar fields transforms as
the scalar in $U_t(P_m,M_{mn})$. It should be noted that the corresponding
finite transformations of $f\star g$  in $U_t(P_m,M_{mn})$ have a form
analogous to the transformation \p{global} of the primary scalar fields
\be
(f\star g)^\prime(x)=(e^{-M_\o}e^{-P_c}f\star g)(x).
\lb{globstar}
\ee
In this pseudolocal representation, we consider the active transformations
of all field objects in the fixed point and do not discuss the dual 
transformations of the noncommutative product of fields together with
the coordinate transformations \footnote{ The action of the dual quantum
Poincar\'e group was discussed in refs.\cite{Oe,KoM}.}. 
 
Applying the map $\mu$ to \p{dtens} yields a noncovariant action of
$U_t(P_m,M_{mn})$ on the commutative product of fields
\bea
&&\hd_\o\star(f g)= -\sfrac12\o^{mn}\mu\circ\D_t(M_{mn})\circ f\otimes g=
(\d_\o f) g+f(\d_\o g)\nn\\
&&+\sfrac{i}2(\o^{mn}\vt_{ns}-\vt^{mn}\o_{ns})\pa_mf \pa^sg\neq 
-\sfrac12\o^{mn}M_{mn}(f g).
\eea 

Four-dimensional integrals of the $\star$-products of fields \p{noncom} are
invariant under the $U_t(P_m,M_{mn})$ symmetry. The quadratic (free)
terms also have the standard Poincar\'e invariance.

\setcounter{equation}0
\section{Deformation of   $N{=}(\sfrac12,\sfrac12)$ supersymmetry }
In this section, we consider the simplest Euclidean supersymmetry 
SUSY$(\sfrac12,\sfrac12)$ generalizing the ISO(4) symmetry of the space
$R^4$. We use the chiral coordinates $z^\M=(y_m, \theta^\a,\bt^\da)$ in
the Euclidean $N{=}(\sfrac12,\sfrac12)$ superspace, where  $m=1,2,3,4,
~\a=1, 2~$ and $\da=\dot{1},\dot{2}$ are the vector and spinor indices of
the group SU(2)$_\sL\times$SU(2)$_\R$. Note that these coordinates are
pseudoreal with respect to  special conjugation \cite{ILZ}
\bea
(y_m)^*=y_m,\q (\th^\a)^*=\ve_\ab \th^\b,\q(\bt^\da)^*=\ve_{\da\db}\bt^\db,\q
(AB)^*=B^*A^* \lb{pseudo}
\eea
for any superfields $A(z)$ and $B(z)$. For instance, one can use the reality
condition for the even Euclidean chiral superfield: $[\phi(y,\th)]^*=
\phi(y,\th)$. 
The central and right  4D coordinates can be expressed via the chiral
coordinates
\be
x_m=y_m-i\th\si_m\bt,\q \bar{y}_m=y_m-2i\th\si_m\bt,
\ee
where $(\si_m)_\ada$ are the Weyl matrices of the group SO(4). Generators
of the supergroup SUSY$(\sfrac12,\sfrac12)$ have the following form:
\bea
&& L^\b_\a=L^\b_\a(y)+L^\b_\a(\th)=\sfrac14(\si_m\bs_n)^\b_\a (y_n\pa_m-
y_m\pa_n)+\th^\b\pa_\a-\sfrac12\d^\b_\a\th^\g\pa_\g,\nn\\
&& R^\db_\da=R^\db_\da(y)+R^\db_\da(\bt)=\sfrac14(\bs_m\si_n)^\db_\da 
(y_m\pa_n-y_n\pa_m)+\bt^\db\bar\pa_\da-\sfrac12\d^\db_\da\bt^\dg\bar\pa_\dg,
\nn\\
&&O=\th^\a\pa_\a-\bt^\da\bar\pa_\da,\q Q_\a=\pa_\a,\q \bar Q_\da=\bar\pa_\da
-2i\th^\a\pa_\ada,\q P_m=\pa_m,\lb{difgen}
\eea
where $(\bs_m)^{\da\a}=\ve^\ab\ve^{\da\db}(\si_m)_\bdb$ 
\footnote{We use the following conventions for the antisymmetric symbols
:\\ 
$\ve_\ab\ve^{\b\g}=\d^\g_\a,\q\ve_{\da\db}\ve^{\db\dg}=\d^\dg_\da$.}, 
and $\pa_\M=(\pa_m, \pa_\a,\bar\pa_\da)$ are the partial derivatives
in the chiral coordinates
\be
\pa_m y_n=\d_{mn},\q\pada=(\si_m)_\ada\pa_m,\q\pa_\a\th^\b=\d^\b_\a,\q
\bar\pa_\da\bt^\db=\d^\db_\da.
\ee
The generators $L^\b_\a, R^\db_\da$ and $O$ correspond to the automorphism
group SU(2)$_\sL\times$SU(2)$_\R\times$O(1,1). 

Let us consider even combinations of the supersymmetry generators and the
corresponding transformation parameters
$c_m, \l^\a_\b, \r^\da_\db, a, \eps^\a$ and $\bep^\da$
\be
P_c=c_mP_m,\q aO,\q L_\l=\l^\a_\b L^\b_\a,\q R_\r=\r^\da_\db R^\db_\da,\q 
Q_\eps=\eps^\a Q_\a,\q\bar Q_\bep=\bep^\da\bar Q_\da.
\ee
In studying the deformations, it is convenient to divide the operators of 
SUSY$(\sfrac12,\sfrac12)$ transformations into two sets 
\bea
&&\d A=(\d_g+\d_G)A=-(g+G)A,\q g=P_c+R_\r+Q_\eps,\q G=L_\l+aO+\bar Q_\bep.
\lb{standtr}
\eea

We let S$(4|2,2)$ and C$(4|2,0)$ denote the standard algebras of general and 
chiral superfields, respectively. The product in these algebras is
supercommutative for superfields $A(z)$ and $B(z)$ with the fixed 
$Z_2$ parities $p(A)$ and $p(B)$
\be
AB=(-1)^{p(A)p(B)}BA.
\ee
This product is defined formally by the bilinear map $\mu$  
\be
\mu\circ A\otimes B=A(z)B(z).
\ee
In  field theory, the tensor product  corresponds to a nonlocal product
of superfields $A(z_1) B(z_2)$ defined on independent sets of coordinates,
and $\mu$ can then be interpreted as providing an identification of $z_1$ and 
$z_2$.

The coproduct is trivial for the generators of SUSY$(\sfrac12,\sfrac12)$  
\be
\D(g)=g\otimes 1+1\otimes g,\q\D(G)=G\otimes 1+1\otimes G.
\ee
This coproduct defines the action of the supersymmetry on the tensor
product of superfields 
\be
\d(A\otimes B)=-\D(g+G)(A\otimes B)=\d A\otimes B+A\otimes \d B
\ee
and yields the standard Leibniz rule for  supersymmetry transformations
on the local product of superfields
\be
\d(AB)=\mu\circ\d(A\otimes B) =(\d A)B+A\d B.
\ee

A non-anticommutative deformation of the coordinates of the Euclidean
$N{=}(\sfrac12,\sfrac12)$ superspace $\hat z=(y_m, \hat\th^\a, \bt^\da)$
was considered in  \cite{Se}. The Clifford coordinate of the deformed
superspace $\hat\th^\a$ satisfies the simple relations
\bea
&&T^\ab=\hat\th^\a\star\hat\th^\b+\hat\th^\b\star\hat\th^\a-C^\ab=0,\lb{basic}
\eea
where $C^\ab$ are some constants, and the coordinates $y_m, \bt^\da$ remain
undeformed
\be
[\hat\th^\a, y_m]=[\bt^\da,y_m]=[y_m,y_n]=0,\q\{\hat\th^\a,\bt^\da\}=\{\bt^\da,
\bt^\db\}=0.
\ee 
The canonical decomposition of the ordered operator superfield $\hat A(\hat z)$
is based on the assumption of antisymmetrization in $\hat\th^\a$ 
\be
\hat A(y,\hat\th,\bt)=r(y,\bt)+\hat\th^\a\chi_\a(y,\bt)+\ve_\ab\hat\th^\a\star
\hat\th^\b s(y,\bt).
\ee 
Non-anticommutative deformations of S$(4|2,2)$ and C$(4|2,0)$ are denoted
by S$_\star(4|2,2)$ and C$_\star(4|2,0)$.

In the pseudolocal representation, the noncommutative superfield $\hat A(\hat z)$ 
corresponds to the supercommutative image $A(z)$ in the undeformed superspace
\be
w[\hat A(y,\hat\th,\bt)]=A(z)=r(y,\bt)+\th^\a\chi_\a(y,\bt)+\ve_\ab\th^\a\th^\b
s(y,\bt).
\ee
The corresponding $\star$-product of  superfields $A(z)$ and $B(z)$ contains
the left supersymmetry generators $Q_\a$
\bea
&&w[\hA\star\hB]=(A\star B)(z)=Ae^P B=A B-\sfrac12(-1)^{p(A)}C^\ab Q_\a A\, 
Q_\b B\nn\\
&&-\sfrac{1}{32}C^\ab C_\ab Q^2A\,Q^2B,\lb{starpr}\\
&&APB=-\sfrac12(-1)^{p(A)}C^\ab Q_\a A\, Q_\b B,\q P^3=0,\nn
\eea 
where  $P$ is the nilpotent bidifferential operator and 
$Q^2=\ve^{\b\a}Q_\a Q_\b$. In the right-hand side of this formula, all
products of superfields and their derivatives are supercommutative. Using
decompositions of these products in powers of $\th^\a$, we can easily
construct the ordered decomposition of the operator product $(\hA\star\hB)(\hat z)
=w^{-1}[(A\star B)(z)]$ in terms of $\hat\th^\a$. Relation  \p{basic}
is satisfied automatically in  representation \p{starpr}.

It is evident that the noncommutative algebra S$_\star(4|2,2)$ is 
noncovariant under transformations of the undeformed supersymmetry
$\d_G$ \p{standtr}, for instance,
\be
\d_\bep (A\star B)=-(\bep^\da \bar Q_\da A)\star B-A\star\bep^\da \bar Q_\da B
\neq -\bep^\da\bar Q_\da(A\star B),
\ee
although the operator  $\d_g$ acts covariantly   
\be
\d_g(A\star B)=(-gA)\star B-A\star(gB)=-g(A\star B). 
\ee

The twist operator  $\cF=\exp(\cP)$ in this superspace was  considered in
\cite{KS} (see, also \cite{Ku,ISa})
\bea
&\cP=-\sfrac12C^\ab Q_\a \otimes Q_\b,\q\cP(A\otimes B)=-\sfrac12
(-1)^{p(A)}C^\ab Q_\a A \otimes Q_\b B,& \nn\\
&\cF(A\otimes B)=A\otimes B -\sfrac12(-1)^{p(A)}C^\ab Q_\a A 
\otimes Q_\b B-\sfrac{1}{32}C^\ab C_\ab Q^2A\otimes Q^2B.& 
\eea
The operator $\cP$ is real with respect to the Hermitian conjugation including
the map \p{pseudo} on the supersymmetry generators $Q_\a^*=Q^\a$ and
the transposition
\be
\cP^*=-\sfrac12(C^\ab)^* Q_\b^* \otimes Q_\a^*=\cP,
\ee
if the condition $(C^\ab)^*=\ve_{\a\r}\ve_{\b\si}C^{\r\si}$ is
satisfied. The reality of the Poisson bracket follows from this property
\be
(APB)^*=\mu\circ \cP^*(B^*\otimes A^*)=B^*PA^*.
\ee

The bilinear map $\mu_\star$ in S$_\star(4|2,2)$ can be defined via the
twist operator
\bea
&&A\star B\equiv\mu_\star\circ A\otimes B=\mu\circ\exp(\cP) A\otimes B.
\eea

By analogy with the map between differential operators on the commutative
and noncommutative algebras of functions \p{diffmap}, the map of the 
$k$-th order differential operator 
$D=\xi^{\M_1\ldots\M_k}(z)\pa_{\M_k}\ldots\pa_{\M_1}$ on the supercommutative
algebra S$(4|2,2)$ to the differential operator $\hX_D$ on S$_\star(4|2,2)$
can be easily defined
\bea
&&(DA)=\mu\circ\xi^{\M_1\ldots\M_k}\otimes\pa_{\M_k}\ldots\pa_{\M_1}A=
\mu_\star\circ\exp(-\cP)\xi^{\M_1\ldots\M_k}\otimes\pa_{\M_k}\ldots\pa_{\M_1}A
\nn\\
&&=\xi^{\M_1\ldots\M_k}\star\pa_{\M_k}\ldots\pa_{\M_1}A+\sfrac12(-1)^{p(D)}
C^\ab Q_\a\xi^{\M_1\ldots\M_k}\star\pa_{\M_k}\ldots\pa_{\M_1}Q_\b A\nn\\
&&-\sfrac{1}{32}C^\ab C_\ab Q^2\xi^{\M_1\ldots\M_k}\star\pa_{\M_k}\ldots
\pa_{\M_1}Q^2 A=(\hX_D\star A),\lb{mapdif}
\eea
where $p(D)$ is the $Z_2$ parity of the operator $D$. The operator $\hX_D$ 
includes derivatives of the orders $k, k+1$ and $k+2$. The differential 
operators on S$(4|2,2)$
form an associative algebra, and the map $D\rightarrow \hX_D$ generates the 
isomorphic algebra of the differential operators on S$_\star(4|2,2)$
\be
(D_1D_2A)=(\hX_{D_1}\star\hX_{D_2}\star A).\lb{isomor}
\ee

In general, the image $\hX_\xi$ for the first-order operator 
$\xi=\xi^\M(z)\pa_\M$ contains  terms with derivatives of the first, second 
and third orders in S$_\star(4|2,2)$
\bea
&&(\hat\pa_\M\star A)=\pa_\M A,\q (\hat\pa_\M\star z^\N)=\d^\N_\M,\nn\\
&&(\hX_{\xi}\star A)=(\xi A)=\mu_\star\circ\exp(-\cP)
(\xi^\M(z)\otimes\pa_\M)(1\otimes A)=\xi^\M(z)\star\pa_\M A\nn\\
&&+\sfrac12(-1)^{p(\xi)}C^\ab Q_\a\xi^\M(z)\star\pa_\M Q_\b A-
\sfrac{1}{32}C^\ab C_\ab Q^2\xi^\M(z)\star\pa_\M Q^2 A.\lb{mapdif1}
\eea 
The noncommutative images of the operators $\bar Q_\da$ and $L^\b_a$ 
\p{difgen} are the second-order differential operators in S$_\star(4|2,2)$:
\bea
&&(\widehat{\bar{Q}}_\da\star A)=(\bar\pa_\da-2i\th^\a\pada+iC^\ab\pada Q_\b)
\star A=\bar Q_\da A,\nn
\\
&&(\hL^\b_\a\star A)=(L^\b_\a-\sfrac12C^{\b\g}Q_\g Q_\a)\star A=L^\b_\a A,
\lb{hatL}
\eea
while the operators $P_m, R^\db_\da$ and $O$ preserve their form under
this map, for instance,
\be
(OA)=(\th^\a\pa_\a-\bt^\da\bar\pa_\da)\star A.
\ee 
The additional term in  $\hat O$ vanishes, $C^\ab Q_\a Q_\b=0$.

Expression \p{mapdif} includes the differentiation and the $\star$-product
in the pseudolocal representation; however,  this formula allows defining
the action of the corresponding operator $(\hX_D\star\hA)=w^{-1}[(\hX_D\star A)]$
on non-anticommutative superfields in an arbitrary representation of the algebra
S$_\star(4|2,2)$. In the operator representation, it is easy to verify that
 the quantity $T^\ab$ \p{basic} is covariant under the action
of the deformed generators
 \bea
&&(\widehat{\bar{Q}}_\da\star T^\ab)=0,\q(\hL^\r_\si\star T^\ab)=\d^\a_\si 
T^{\r\b}+\d^\b_\si T^{\a\r}-\d^\r_\si T^{\a\b}.
\eea 

According to relation \p{isomor}, the deformed generators on S$_\star(4|2,2)$
form the Lie superalgebra isomorphic to the undeformed Lie superalgebra of
generators \p{difgen}.

The coproduct $\D_t(G)=e^{-\cP}\D(G)e^\cP$ in the deformed supersymmetry
SUSY$_t(\sfrac12,\sfrac12)$ is changed on some of the  generators
\bea
&\D_t(\bar{Q}_\bep)
=\bar{Q}_\bep\otimes 1+1\otimes \bar{Q}_\bep+i\bep^\da C^\ab(\pa_\ada\otimes 
Q_\b-Q_\a\otimes \pa_\bda),&\lb{coprbQ}\\
&\D_t(L_\l)=L_\l\otimes 1+1\otimes L_\l
+\sfrac12C^{\r\si}(\l^\a_\r Q_\a\otimes Q_\si+\l^\a_\si Q_\r\otimes Q_\a),&
\lb{coprL}\\
&\D_t(O)=O\otimes 1+1\otimes O-C^\ab Q_\a\otimes Q_\b.&\lb{coprO}
\eea
The coproduct is not deformed on the generators $R^\db_\da, P_m$ and $  Q_\a$.

By definition, the action of SUSY$_t(\sfrac12,\sfrac12)$ on a primary
superfield $\hA$ in an arbitrary representation is generated by the
undeformed supersymmetry transformations of the supercommutative image
$A(z)$
\bea
&&\hd\star A=-(g+G)A=-(\hat g+\hG)\star A,
\lb{primar}\\
&&\hd\star \hA=w^{-1}[\hd\star A]=-(\hat g+\hG)\star\hA,\nn
\eea
where relations between operators and superfields in different representations
are used.

The deformed coproduct $\D_t(G)$ in SUSY$_t(\sfrac12,\sfrac12)$ determines
transformations of the noncommutative product in the algebra S$_\star(4|2,2)$
\bea
&&\hd_\bep\star(A\star B)=-\mu_\star\circ\D_t(\bar{Q}_\bep) A\otimes B=
-(\bar{Q}_\bep A)\star B-A\star\bar{Q}_\bep B\nn\\
&&-i\bep^\da C^\ab[(-1)^{p(A)}\pa_\ada A\star Q_\b B-Q_\a A\star \pa_\bda B],
\lb{beptr}\\
&&\hd_\l\star(A\star B)=-\mu_\star\circ\D_t(L_\l) A\otimes B=-(L_\l A)\star B
-A\star L_\l B\nn\\
&&-\sfrac12(-1)^{p(A)}C^{\r\si}(\l^\a_\r Q_\a A\star Q_\si B+\l^\a_\si Q_\r 
A\star Q_\a B),\lb{Ltr}\\
&&\hd_a\star(A\star B)=-a(OA)\star B-aA\star (OB)+a(-1)^{p(A)}C^\ab Q_\a 
A\star Q_\b B.
\eea

The appearance of terms with $C^\ab$ in the transformations of $A\star B$ can be
treated as a deformation of the Leibniz rules for $\hd_\bep, \hd_\l$ and $\hd_a$. 
It is not difficult to show that these relations yield the covariance of the 
noncommutative product 
\be
\hd_G\star (A\star B)=-G(A\star B),
\ee
which transforms similarly to primary superfields $A$ and $B$ \p{primar} 
in SUSY$_t(\sfrac12,\sfrac12)$. For instance, the formula
$\hd_\bep\,\star(A\star B)=-\bar Q_\bep(A\star B)$ is derived from eq.\p{beptr}
using the  relations
\bea
&&-\bar Q_\bep(A\star B)=-[\bar Q_\bep,(Ae^PB)]=-(\bar Q_\bep A)e^PB
-A[\bar Q_\bep,e^P]B-Ae^P(\bar Q_\bep B),\nn\\
&&-A[\bar Q_\bep,e^P]B=-i\bep^\da C^\ab[(-1)^{p(A)}\pa_\ada A\star Q_\b B
-Q_\a A\star \pa_\bda B].
\eea 

We note that superfield $AB$ is a noncovariant quantity in SUSY$_t(\sfrac12,
\sfrac12)$. For example, it is easy to define noncovariant actions of the
operators 
$\hd_\bep$ and $\hd_\l$ on the ordinary product of even chiral superfields
\bea
&&\hd_\bep\star(\phi_1 \phi_2)\equiv -\mu\circ\D_t(\bar{Q}_\bep)\phi_1\otimes
\phi_2 = -\bar Q_\bep (\phi_1\phi_2)-i\bep^\da C^\ab(\pa_\ada \phi_1\, Q_\b 
\phi_2\nn\\
&&-Q_\a \phi_1\, \pa_\bda \phi_2)=\hd_\bep(a_1a_2)+\th^\a\hd_\bep(a_1\psi_{2\a}
+a_2\psi_{1\a})+O(\th^2),\nn\\
&&\hd_\l\star(\phi_1 \phi_2)\equiv -\mu\circ\D_t(L_\l)\phi_1\otimes\phi_2=
-\l^\a_\b L^\b_\a(\phi_1 \phi_2)\nn\\
&&-\sfrac12 C^{\r\si}(\l^\a_\r Q_\a\phi_1 Q_\si\phi_2+\l^\a_\si Q_\r\phi_1 Q_\a
\phi_2)=\hd_\l(a_1a_2)+O(\th).\lb{noncov1}
\eea
The first terms in these formulas coincide with the transformations of the
undeformed supersymmetry. Using the  $\th$-decomposition of these superfield
formulas
 \bea
&&\phi_i=a_i+\th^\a\psi_{i\a}+\th^2f_i,\\
&&Q_\a \phi_i=\psi_{i\a}+2\th_\a f_i,\q Q^2\phi_i=-4f_i\nn
\eea
one can obtain the deformed transformations of component fields, for instance, 
\bea
&& \hd_\bep(a_1a_2)=-i\bep^\da C^\ab(\pada a_1\,\psi_{\b 2}-\psi_{\a 1}
\pa_{\b\da}a_2),\nn\\
&&\hd_\l(a_1a_2)=-\l^\a_\b L^\b_\a(y)(a_1a_2)-\sfrac12C^{\r\si}(\l^\a_\r 
\psi_{\a 1} \psi_{\si 2}+\l^\a_\si \psi_{\r 1} \psi_{\a 2}) .\lb{noncov2}
\eea
 
The expansion of the $\star$-product of two chiral superfields in $\th^\a$
depends on the constants $C^\ab~$
\bea
&&\Phi_{12}=\phi_1\star\phi_2=\cB+\th^\a\Psi_\a +\th^2F,\lb{defcomp}\\
&&\cB=a_1a_2-\sfrac12C^\ab\psi_{1\a}\psi_{2\b}-\sfrac12C^\ab C_\ab f_1f_2,
\nn\\
&&\Psi_\a=a_1\psi_{2\a}+a_2\psi_{1\a}-C_\ab(f_1\psi_2^\b-f_2\psi_1^\b),\nn\\
&&F=a_1f_2+a_2f_1-\sfrac12\psi^\a_1\psi_{2\a}.
\eea
These relations generate the deformed tensor calculus for the product of
the chiral component multiplets. The transformations of the composed components
\p{defcomp} in the deformed supersymmetry are completely analogous to the
transformations of the primary component fields $a_i, \psi_{\a i}$ and $ f_i$
\bea
&&\hd_\bep \cB=0,\q \hd_\bep\Psi_\a=-2i\bep^\da\pada\cB,\q
\hd_\bep F=-i\bep^\da\pada\Psi^\a,\nn\\
&&\hd_\l\cB=-\l^\a_\b L^\b_\a(y)\cB,\q \hd_\l\Psi_\g=\l^\a_\g\Psi_\a-
\l^\a_\b L^\b_\a(y)\Psi_\g,\q
\hd_\l F=-\l^\a_\b L^\b_\a(y)F .\lb{compcov}
\eea
These transformations are compatible with the noncovariant transformations
of the products of components \p{noncov2}.

The non-anticommutative deformation of the Euclidean model with an
arbitrary number of chiral (antichiral) superfields and gauge superfields
$V(z)$ involves $\star$-products of these superfields in the superfield
action \cite{Se}. Each term of the $\star$-polynomial decomposition
of this action is separately invariant with respect to the transformations
of SUSY$_t(\sfrac12,\sfrac12)$, and the quadratic terms are also invariant
under the ordinary supersymmetry.

\setcounter{equation}0
\section{Deformed  $N{=}(1,1)$ supersymmetry } 
Nilpotent deformations of the Euclidean $N{=}(1,1)$ supersymmetry were
considered in the framework of the harmonic-superspace formalism \cite{ILZ,FS}
using the SU(2)/U(1) harmonics $u^\pm_i$ and the chiral superspace coordinates
\be
z^\M=(y_m, \th^\a_k, \bt^{\da k})\, , \q y_m = x_m+
i\th_k\si_m\bt^k,
\ee
where $x_m$ are the central 4D coordinates. Standard conjugation of these
coordinates changes positions of all spinor indices
\be
\overline{y_m}=y_m,\q \overline{\th^\a_k}= \th_\a^k,\q \overline{\bt^{\da k}}
=-\bt_{\da k}
\lb{standconj} 
\ee
and  in particular preserves the invariance under the $SU(2)$ automorphisms
acting on index $k$. In the same superspace, the alternative
pseudoconjugation can be defined as \cite{ILZ}
\be
(y_m)^*=y_m,\q (\th^\a_k)^*= \th_{\a k},\q (\bt^{\da k})^*=\bt_\da^k,
\lb{pseudo2}
\ee
which does not change the position of the $SU(2)$-index $k$. The spinor 
derivatives $D^k_\a$ and $\bar D_{\da k}$ in these coordinates are given by
\be
D^k_\a=\pa^k_\a+2i\bt^{\da k}\pada,\q\bar D_{\da k}=\bar\pa_{\da k}.
\ee

The even part of the harmonic superspace $R^4\times S^2$ has a dimension
4+2; it is convenient to use separate symbols for the left and right odd dimensions
of the Grassmann coordinates of the general superspace (4,4), the chiral superspace
(4,0), and the analytic superspace (2,2), respectively. We  use symbol
S$(4,2|4,4)$ for the supercommutative algebra of general harmonic superfields
and S$_\star(4,2|4,4)$ for the corresponding non-anticommutative algebra.
 
We use the following representation of the  SUSY(1,1) 
supersymmetry generators as the differential operators on the algebra S$(4,2|4,4)$:
\bea
&&T_l^k=-\th^\a_l\pa^k_\a+\sfrac12\d^k_l\th^\a_j\pa^j_\a+\bt^{\da k}
\bar\pa_{\da l}-\sfrac12\d^k_l\bt^{\da j}\bar\pa_{\da j}-u^\pm_l\pa^{\mp k}
+\sfrac12\d^k_l u^\pm_j\pa^{\mp j},\nn\\
&&L_\a^\b=L^\b_\a(y)+\th^\b_k\pa_\a^k-\sfrac12\d^\b_\a\th^\g_k\pa_\g^k,\q 
R^\db_\da=R^\db_\da(y)+\bt^{\db k}\bar\pa_{\da k}-\sfrac12\d^\db_\da
\bt^{\dg k}\bar\pa_{\dg k},\nn\\
&& O=\th^\a_k\pa_\a^k-\bt^{\da k}\bar\pa_{\da k},\q Q_\a^k=\pa^k_\a,\q 
\bar Q_{\da k}=\bar\pa_{\da k}-2i\th^\a_k\pada,\q P_m=\pa_m,\lb{dif11}
\eea
where $L^\b_\a(y)$ and $R^\db_\da(y)$ are defined in \p{difgen} and the
partial derivatives satisfy the relations
\be
\pa_m y_n=\d_{mn},\q \pa^i_\a\th^\b_k=\d^i_k\d^b_\a,\q
\bar\pa_{\da i}\bt^{\db k}=\d^k_i\d^\db_\da,\q\pa^{\mp l}u^\pm_k=\d^l_k,.
\ee

To study deformations, it is convenient to separate the SUSY(1,1) 
transformations into two parts
\be
\d_gA=-gA,\q\d_G A=-GA,
\ee
using the following combinations of generators and the corresponding
parameters:
\bea
&&g=P_c+R_\r+Q_\eps,\q G=T_u+L_\l+\bar Q_\bep+aO,\lb{sep11}\\
&&P_c=c_mP_m,\q T_u=u^k_lT^l_k,\q L_\l=\l^\a_\b L^\b_\a,\q R_\r=\r^\da_\db 
R^\db_\da,\q Q_\eps=\eps^\a_kQ^k_\a,\q \bar Q_\bep=\bep^{\da k}\bar Q_{\da k}.
\nn
\eea

The analytic coordinates of the harmonic superspace $(x_A, \th^\pm, 
\bt^\pm)$  can be defined using the harmonic projections
of the Grassmann coordinates $\th^{\pm\a}=u^\pm_k\th^{\a k}$ and 
$\bt^{\pm\da}=u^\pm_k\bt^{\da k}$. The corresponding representation of the
spinor and harmonic derivatives can be found in  \cite{ILZ,FILSZ}
\bea
&&\Dpa =\partial_{-\alpha} \ ,\qq
\Dma =-\partial_{+\alpha} + 2\ii\btma\partial_{\alpha\dot\alpha}\ ,\nn\\ 
&&\bDpa =\bar\partial_{-\da} \ ,\qq
\bDma =-\bar\partial_{+\da} - 2\ii\tma\partial_{\alpha\dot\alpha}\ ,
\lb{Aspder}\\
&& \Dp_\A=\dpp -2 \ii\tpa \btpa\partial_{\alpha\dot\alpha}
+ \tpa \partial_{-\alpha} + \btpa\bar\partial_{-\da}
\lb{Aharder}
\eea
where $\partial_{\pm\alpha}\equiv\partial/\partial\theta^{\pm\alpha}$,
$\bar\partial_{\pm\da}\equiv\partial/\partial\bar\theta^{\pm\da}$.

The $N{=}(1,1)$ twist operator $\cF=\exp{(\cP)}$ contains the nilpotent
operator
\be
\cP=-\sfrac12C^\ab_{kl}Q^k_\a\otimes Q^l_\b,\q\cP^5=0,\lb{PQdef}
\ee
where $C^\ab_{kl}$ are the deformation constants. The operator $\cP$ is Hermitian
under the conjugation constructed from the  transposition and
conjugation \p{standconj}
\be
\overline\cP=-\sfrac12\overline{C^\ab_{kl}}\overline{Q^l_\b}\otimes
\overline{Q^k_\a}=\cP,\q\overline{Q^k_\a}=\ve^{\a\r}\ve_{kj}Q^j_\r,
\ee
if the conditions $\overline{C^\ab_{kl}}=\ve_{\a\r}\ve_{\b\si}
\ve^{ki}\ve^{lj}C^{\r\si}_{ij}$ are satisfied.

The action of the operator $\cP$ on the tensor product of superfields $A$ 
and $B$ is compatible with the $Z_2$ grading
\be
\cP A\otimes B= -\sfrac12(-1)^{p(A)}C^\ab_{kl}Q^k_\a A\otimes Q^l_\b B.
\ee
A non-anticommutative product in the corresponding deformed algebra 
S$_\star(4,2|4,4)$ can be defined using the equivalent
formulas
\bea
&&A\star B=A\exp(P)B=\mu\circ\exp{(\cP)}A\otimes B=\mu_\star\circ A\otimes B
\eea
where $\mu$ and $\mu_\star$ are the bilinear maps for S$(4,2|4,4)$ and
S$_\star(4,2|4,4)$, and $P$ is the bidifferential operator in \cite{ILZ,FS}
\be
A PB=-\sfrac12(-1)^{p(A)}C^\ab_{kl}Q^k_\a A\, Q^l_\b B=\mu\circ\cP\,A\otimes B.
\lb{bidiff}
\ee

The non-anticommutative algebras of the $N{=}(1,1)$ chiral or analytic superfields
are defined as subalgebras of S$_\star(4,2|4,4)$ using the superfield
constraints
\be
\bar D_{\da k}B=0,\q\mbox{or}~(D^+_\a,\bDpa)\L=0,
\ee
which are preserved by the deformation operator $\exp(P)$.

By analogy with eq. \p{mapdif}, each differential operator on the algebra
S$(4,2|4,4)$ corresponds to the operator on the noncommutative
algebra S$_\star(4,2|4,4)$, for instance,
 \bea
&& \widehat{\bar Q}_{\da k}\star A=\bar Q_{\da k} A,\nn\\
&&\widehat{\bar Q}_{\da k}=\bar\pa_{\da k}-2i\th^\a_k\star\pada-C^\ab_{kl}
\pa^l_\b\pada.
\eea
In the twist-deformed supersymmetry SUSY$_t$(1,1), we can use the standard
representation of generators \p{dif11} on the primary superfields.
 
The deformed coproduct in   SUSY$_t$(1,1), $\D_t(\cG)=e^{-\cP} \D(\cG)
e^\cP$, can be easily calculated on the following generators:
\bea
&&\D_t(\bar Q_\bep)=\bar Q_\bep\otimes 1+1\otimes \bar Q_\bep+i\bep^{\da k}
C^\ab_{kj}\pada\otimes Q^j_\b-i\bep^{\da k}C^\ab_{ik}Q^i_\a\otimes  
\pa_{\b\da},\nn\\
&&\D_t(T_u)=T_u\otimes 1+1\otimes T_u-\sfrac12u^l_kC^\ab_{lj}Q^k_\a\otimes 
Q^j_\b-\sfrac12u^l_kC^\ab_{jl}Q^j_\a\otimes  Q^k_\b,
\nn\\
&&\D_t(L_\l)=L_\l\otimes 1+1\otimes L_\l+\sfrac12\l^\a_\b C^{\b\g}_{kl} Q^k_\a
\otimes Q^l_\g+\sfrac12\l^\a_\b C^{\r\b}_{kl} Q^k_\r\otimes  Q^l_\a,\nn\\
&&\D_t( O)=O\otimes 1+1\otimes O-C^\ab_{kl}Q^k_\a\otimes Q^l_\b.\lb{copr11}
\eea
The deformation of the coproduct vanishes for the operator $g=Q_\eps+P_c+R_\r$: 
$e^{-\cP}\D(g)e^\cP=1\otimes g+g\otimes 1$. 

The noncommutative  $\star$-products of the  $N{=}(1,1)$ superfields preserve
the covariance under the deformed transformations of SUSY$_t$(1,1)
\bea
&&\hd_G\star(A\star B)=- \mu_\star\circ \D_t(G)A\otimes B=-G(A\star B).
\eea

The deformed Leibniz rules for the $\star$-products are derived from formulas
\p{copr11}, for instance,
\bea
&&\hd_\bep\star(A\star B)=-\mu_\star\circ\D_t(\bar Q_\bep)A\otimes B=
-(\bar Q_\bep A)\star B+A\star (\bar Q_\bep B)\nn\\
&&-i(-1)^{p(A)}C^\ab_{kj}\bep^{\da k}\pada A
\star Q^j_\b B+iC^\ab_{ik}\bep^{\da k}Q^i_\a A\star \pa_{\b\da}B=
-\bar Q_\bep(A\star B).
\eea

The transformation $\hd_\bep$ acts noncovariantly on the supercommutative
product of superfields
\bea
&&\hd_\bep\star(A B)=-\mu\circ\D_t(\bar Q_\bep)A\otimes B=-\bar Q_\bep (A B)
\nn\\
&&-i(-1)^{p(A)}C^\ab_{kj}\bep^{\da k}\pada A Q^j_\b B+iC^\ab_{ik}\bep^{\da k}
Q^i_\a A \pa_{\b\da}.
\eea
It is easy to define the deformation of transformations for products
of the $N{=}(1,1)$ component fields using the corresponding Grassmann
expansions of the superfield transformations.

In the special case of the singlet deformation \cite{FILSZ,ILZ2}, the twist
operator contains the parameter $I$ and the SU(2)$\times$SU(2)$_\sL$ invariant
constant tensor
\be
C^\ab_{kl}=2I\ve^\ab\ve_{kl}~\Rightarrow~\cP_s=-IQ^i_\a\otimes Q_i^\a .
\ee
The deformation corresponding to the operator $\cP_s$ vanishes on the generators of
the SU(2) and SU(2)$_\sL$ transformations and remains only for the
$N=(0,1)$ and O(1,1) generators
\bea
&&\D_t(\bar Q_{\da k})=\bar Q_{\da k}\otimes 1+1\otimes \bar Q_{\da k}+2iI\pada
\otimes Q_k^\a-2iIQ_k^\a\otimes  \pa_{\a\da},\nn\\
&&\D_t( O)=O\otimes 1+1\otimes O-2IQ^k_\a\otimes Q_k^\a.\lb{coprPs}
\eea

The degenerate $N{=}(1,\sfrac12)$ deformation in  \cite{ILZ} corresponds
to  using  the twist operator $\cF_{deg}=\exp(\cP_{deg})$, where the operator
\be
\cP_{deg}=-\sfrac12C^\ab Q^2_\a\otimes Q^2_\b
\ee
is Hermitian under the alternative pseudoconjugation \p{pseudo2}
\be
(Q^k_\a)^*=Q^{\a k},\q (\bar Q^{\da k})^*=\bar Q^k_\da,\q (C^\ab)^*=C_\ab.
\ee
In this case, the coproduct $\D_t$ is deformed on the generators 
$\bar Q_{\da 2}, L^\b_\a, O$ and $T^l_k$.

In \cite{IZ}, we considered the deformation of S$(4,2|4,4)$ acting
simultaneously in the chiral and antichiral sectors of the superspace
and using the pseudoconjugation \p{pseudo2}. This deformation corresponds to
the twist operator $\widehat{\cF}=\exp(\widehat{\cP})$
\be
\widehat{\cP}=-\sfrac12C^\ab Q^2_\a\otimes Q^2_\b-\sfrac12\bar{C}^{\da\db}
\bar Q_{\da 1}\otimes \bar Q_{\db 1}-B^\ada(Q^2_\a\otimes\bar Q_{\da 1}
+\bar Q_{\da 1}\otimes Q^2_\a).\lb{hatdef}
\ee 
It is interesting  that an analogous twist operator can be
used to deform  the $N{=}2, D{=}(3,1)$ superspace on the basis of
the Minkowski space. The Hermitian symmetry of $\widehat{\cP}$ is possible in 
this case if the alternative conjugation is used in the central coordinates 
of the $N{=}2, D{=}(3,1)$ superspace
\be
(x^m)^\dagger=x^m,\q (\th^\a_k)^\dagger=\bt^\da_k,\q(\bt^\da_k)^\dagger=
\th^\a_k,\lb{altconj}
\ee
which breaks down the automorphism group SU(2) but is compatible with
covariant conjugation of spinors in the group SL(2,C)
\footnote{We note that the alternative and usual conjugations
act identically on the SU(2) invariant quantities.}. Operator \p{hatdef}
deforms the coproduct on the generators of SL(2,C), U(2), 
$Q^1_\a$ and $ \bar Q_{\da 2}$.

The differential operators $(\pa_m, D^k_\a,\bar D_{\da k},\pa/\pa u^\pm_k)$   
satisfy the standard Leibniz rules for all nilpotent deformations considered.

We consider the primary analytic superfields of the hypermultiplet
$q^+,~ \tilde q^+$ and the gauge multiplet $V^{++}$ \cite{ILZ,FS}, which
have the following noncommutative gauge transformations:
\bea
&&\delta_\Lambda V^{++}=D^{++}\Lambda+[V^{++}, \Lambda]_\star\,,\q
\delta_\L q^+=[q^+,\L]_\star.
\lb{GaugetranV}
\eea
where $\L$ is the superfield gauge parameter. These superfields and their
$\star$-products transform covariantly in the deformed supersymmetry 
SUSY$_t$(1,1)
\bea
&&\hd_G\star[V^{++}(z,u_1)\star V^{++}(z,u_2)]=-G [V^{++}(z,u_1)\star 
V^{++}(z,u_2)],\\
&&\hd_G\star(V^{++}\star q^+) =-G(V^{++}\star q^+),
\eea
where the generators $G$ are given by eq. \p{sep11} or by the equivalent
relations in the analytic coordinates.

The gauge action of $V^{++}$ is defined in the full or chiral superspaces
\cite{ILZ,FS,FILSZ}. In an arbitrary gauge, this action is invariant 
under the SUSY$_t$(1,1) transformations. In the analytic superspace,
the superfield action of the hypermultiplets contains the integral with
the measure  $ d^4x_\A (D^-)^4$, and the simple example of the analytic
density has the  form 
 \be
L^{+4}_\star=\tilde q^+\star(\Dp q^+ +[\Vp,q^+]_\star)+
\l q^+\star q^+\star\tilde q^+\star\tilde q^+.
\ee
The deformed transformations of this superfield density are covariant
in an arbitrary gauge of $V^{++}$
\bea
&&(\hd_\bep+\hd_u+\hd_l+\hd_a)\star L^{+4}_\star=(\bep^{\da k}\bar Q_{\da k}
+u^l_kT^k_l+l^\a_\b L^\b_a+aO)L^{+4}_\star,
\eea
and the analytic-superspace integral of these variations vanishes. 
All superfield actions using the $\star$-products in the non-anticommutative
harmonic superspace \cite{FILSZ,ILZ2,CILQ} are invariant with respect to the 
transformations of the deformed supersymmetry SUSY$_t$(1,1). The free quadratic
terms of these theories also preserve  the undeformed  $N=(1,1)$ supersymmetry.
 
The gauge superfield in the WZ-gauge defines the component fields of the
vector multiplet 
\bea
&&V^{++}_{\W\Z}= (\tp)^2\bph(x_\A)+(\btp)^2\phi(x_\A)+2(\tp\si_m\btp)
A_m(x_\A) + 4(\btp)^2\theta^{+\a} u^-_k\Psi^k_\a(x_\A)\nn\\
&&+ \,4(\tp)^2\btp_\da u^-_k\bar\Psi^{\da k}(x_\A)
+3(\tp)^2(\btp)^2u^-_ku^-_l D^{kl}(x_\A).\lb{WZanal}
\eea
The SUSY$_t$(1,1) transformations of the quantity $V^{++}_{\W\Z}$ contain
the standard terms with the supersymmetry generators \p{sep11} complemented
by the composite gauge transformations
\be
(\hd_\eps+\hd_\bep)\star V^{++}_{\W\Z}=-(Q_\eps+\bar Q_\bep) V^{++}_{\W\Z}-
D^{++}(\Lambda_\eps+\L_\bep)-[V^{++}, (\L_\eps+\L_\bep)]_\star,\lb{compos}
\ee
where the $N{=}(1,1)$ supersymmetry generators are considered in the
analytic coordinates
\bea
&& Q^k_\a=-u^{+k}\pa_{+\a}-u^{-k}\pa_{-\a}+2iu^{-k}\btpa\pada,\nn\\
&&\bar Q_{\da k}=u^+_k\bar\pa_{+\da}+u^-_k\bar\pa_{-\da}+2iu^-_k\tpa\pada.
\eea

The composite parameters of the $N{=}(1,1)$ transformations in the  WZ-gauge
are given by
\bea
\L_\eps=2\eps^{-\a}[\tp_\a\bph+\btpa A_\ada+u^-_l(\btp)^2\J^l_\a
+2u^-_l\tp_\a\btp_\da \bJ^{\da l}+u^-_lu^-_j\tp_\a(\btp)^2D^{lj}],\nn&&\\
\L_\bep=2\bep^{-\da}[\btp_\da\phi+\tpa A_\ada+u^-_l(\tp)^2\bJ^l_\da
+2u^-_l\btp_\da\tpa \J_\a^l+u^-_lu^-_j\btp_\da(\tp)^2D^{lj}],&&
\eea
where $\eps^{-\a}=\eps^{\a k}u^-_k$ ¨ $\bep^{-\da}=\bep^{\da k}u^-_k$. The
deformed transformations of the vector-multiplet components are determined
from the Grassmann expansion of the transformations of  $V^{++}_{\W\Z}$ in
\p{compos}. These transformations contain nonlinear gauge terms. The
SUSY$_t$(1,1) transformations for the hypermultiplets are compatible with the
transformation of $V^{++}_{\W\Z}$
\be
(\hd_\eps+\hd_\bep)\star q^+=-(Q_\eps+\bar Q_\bep)q^+ +[q^+,(\L_\eps+
\L_\bep)]_\star.
\ee
Additional terms with $\Lambda_\eps$ and $\Lambda_\bep$ do not violate
the SUSY$_t$(1,1) invariance of the action by virtue of the gauge symmetry.

The singlet  $D$-deformation of the $N{=}(1,1)$ supersymmetry was  considered
in  \cite{ILZ,FS}. This deformation corresponds to the alternative singlet
twist operator constructed using the spinor derivatives $D^k_\a$
\be
(A\star B)_D=\mu\circ \exp(\cP_D) A\otimes B,\q\cP_D=-JD^k_\a\otimes D^\a_k,
\lb{PDsing}
\ee
where $J$ is some constant.  The Leibniz rules for the noncommutative
product are now deformed for the operator $\bar D_{\da k}$ 
\bea
&&\bar D_{\da k}(A\star B)_D=(\bar D_{\da k}A\star B)_D+(-1)^{p(A)}(A\star
\bar D_{\da k} B)_D\nn\\
&&-2iJ(-1)^{p(A)}\pada A\star D^\a_kB-2iJD^\a_kA\star\pada B
\eea
and for the O(1,1) generator, but this deformation vanishes for
the remaining generators of SUSY(1,1) and the spinor derivative $D^k_\a$.

The singlet operator $\cP_D$  \p{PDsing} does not preserve chirality but
does preserve anti-chirality and  Grassmann analyticity. The $D$-deformation
is interesting for the general objects of the superfield geometry in the
$N{=}=(1,1)$ gauge theory \cite{FS}.

\section{ Conclusions}
We have analyzed the $t$-deformations of the Euclidean $N{=}(\sfrac12,
\sfrac12)$ and $N{=}(1,1)$ supersymmetries using the twist operators depending
on the left supersymmetry generators. In this approach, the coproduct in the
universal enveloping supersymmetry algebra is deformed, and the Lie superalgebra
remains undeformed. In the pseudolocal representation, the deformed supersymmetry
is covariantly realized on the noncommutative $\star$-products of  primary
fields, while the $t$-supersymmetry transformations of the supercommutative
product of superfields depend on the deformation parameters. A map of the
differential operators on the ordinary superspace to the differential
operators in an arbitrary representation of the deformed noncommutative 
superfields was constructed. In this representation, the part of the 
$t$-supersymmetry generators is realized by the second-order differential
operators, and it therefore becomes evident that the corresponding 
transformations of the $\star$-product do not satisfy the usual Leibniz rule.
    
The covariance of the $\star$-product and the invariance of the superfield
action using this product are the main principles of the superfield formalism
of the $t$-deformed theories. The bilinear free terms of the deformed action
are also invariant under the usual supersymmetry. The deformation constants
of the non-anticommutative superfield theories violate some initial
(super)symmetries; however, these constants are compatible with the deformed
supersymmetries.  The invariance of the superfield formalism under
the $t$-supersymmetry should be used to study the renormalizability of the 
deformed supersymmetric theories. The formalism of deformations can be
regarded as an interesting analog of the spontaneous
(super)symmetry breaking mechanism if the deformation does not destroy the 
good quantum properties of the supersymmetric field theories.
   
The author is grateful to Professors J. Jost and D. Leites for the kind
hospitality in Institute for Mathematics in Sciences where the part of this 
work was made. This work is partially supported by  grant RFBR 06-02-16684, 
by  grant DFG 36 RUS 113/669-3,  by  NATO grant 
PST.GLG.980302 and grant of the Heisenberg-Landau programme.

\end{document}